\documentclass[largeformat]{interact}
\usepackage[utf8]{inputenc}

\usepackage{epstopdf}
\usepackage[caption=false]{subfig}
\usepackage[sort&compress,comma,super]{natbib}
\usepackage{amsfonts}
\usepackage{booktabs}
\usepackage{tensor}
\usepackage{xcolor}
\usepackage{multirow}
\usepackage{url}
\bibpunct[, ]{[}{]}{,}{n}{,}{,}
\renewcommand\bibfont{\fontsize{10}{12}\selectfont}

\theoremstyle{plain}

\theoremstyle{definition}

\theoremstyle{remark}

\newcommand{\kk}{\hat{\kappa}}
\newcommand{\EE}{\hat{E}^{-}}
\newcommand{\Ham}{\mathcal{H}}
\newcommand{\m}{\mu}
\newcommand{\n}{\nu}
\newcommand{\s}{\sigma}

\begin{document}

\title{A black-box, general purpose quadratic self-consistent field code with and without Cholesky Decomposition of the two-electron integrals}

\author{
\name{Tommaso Nottoli\textsuperscript{a}, J\"urgen Gauss\textsuperscript{b} and Filippo Lipparini\textsuperscript{a}\thanks{Email: filippo.lipparini@unipi.it} }
\affil{\textsuperscript{a}Dipartimento di Chimica e Chimica Industriale, Universit\`a di Pisa, Via G. Moruzzi 13, I-56124 Pisa, Italy; \textsuperscript{b}Department Chemie, Johannes Gutenberg-Universit\"at Mainz, Duesbergweg 10-14, D-55128 Mainz, Germany}
}

\maketitle

\begin{abstract}
We present the implementation of a quadratically convergent Self-consistent field (QCSCF) algorithm based on an adaptive trust-radius optimization scheme for restricted open-shell Hartree-Fock (ROHF), restricted Hartree-Fock (RHF), and  unrestricted Hartree-Fock (UHF) references. The algorithm can exploit Cholesky decomposition (CD) of the two-electron integrals to allow calculations on larger systems. The most important feature of the QCSCF code lies in its black-box nature -- probably the most important quality desired by a generic user. As shown for pilot applications, it does not require one to tune the self-consistent field (SCF) parameters (damping, Pulay’s DIIS, and other similar techniques) in difficult-to-converge molecules. Also, it can be used to obtain a very thigh convergence with extended basis set -- a situation often needed when computing high-order molecular properties -- where the standard SCF algorithm starts to oscillate. Nevertheless, trouble may appear even with a QCSCF solver. In this respect, we discuss what can go wrong, focusing on the multiple UHF solutions of ortho-benzyne. 
\end{abstract}


\section{Introduction}
Self-consistent field (SCF) methods are the starting point of virtually every quantum chemistry application. Kohn-Sham (KS) density-functional theory\cite{Kohn65} (DFT), Hartree-Fock\cite{Hartree1957} (HF) and even semiempirical methods\cite{Thiel2000} require as a fundamental numerical step the variational optimization of the energy with respect to the orbitals, that are determined in the process. The standard algorithm to solve the optimization problem -- the self-consistent field algorithm itself\cite{Roothan51,Roothaan1960} -- is the iterative solution of a non-linear eigenvalue problem, which is solved using a fixed-point approach. As the problem is strongly non-linear, a simple iterative procedure is often not sufficient to achieve convergence. In the last several decades, various strategies have been developed to improve the reliability and stability of the SCF algorithm\cite{Rabuck1999,Cances2000}, most notably Pulay's direct inversion in the iterative subspace\cite{Pulay1980,Pulay1982,Hamilton1985} (DIIS) and various refinement thereof\cite{Sellers1993,Cances2000,Kudin2002}, damping of the SCF iterations\cite{Karlstrom1979}, level shifting\cite{Saunders1973}, and combinations of the methods. Advanced strategies to compute accurate guesses of the initial density have also greatly improved the overall reliability of the method\cite{Vacek1999,Lethola19}, to the point that closed-shell systems seldom pose convergence problems in real life applications.

Unfortunately, things are not so easy when one is dealing with open-shell systems, or even with closed-shell systems with small energy gaps between the highest occupied and lowest unoccupied molecular orbital (HOMO, LUMO). Furthermore, even in well-behaved cases, achieving a very tight convergence of the SCF density can be difficult, but mandatory for applications involving high-order response properties or geometrical derivatives, especially if a post-HF method is used. Being frustrated by problematic SCF convergence is therefore a common experience for computational chemists. For these reasons, alternative strategies and robust numerical procedures are still object of active investigation, despite SCF being possibly the most well-established technique in computational chemistry. 

An alternative strategy to solving the SCF problem is to use a standard optimization technique, and, in particular, a second-order method\cite{Bacskay1981,Bacskay1982}. In second-order methods, the SCF energy is parametrized as a function of non-redundant orbital rotations and expanded up to second order in a Taylor series, obtaining thus a quadratic energy model. The latter is then optimized to find a step and the process is iterated until convergence. The straightforward Newton-Raphson (NR) method just described suffers however of a small convergence radius, which can cause an erratic or even divergent behavior if the optimization process is started far from a local minimum. It is possible to constrain the minimization so that the computed step is no larger than a user-defined trust radius. The trust-radius Newton method, known as Levenberg-Marquardt\cite{Fletcher1999} (LM) optimization, can be further coupled with an adaptive choice of the trust radius, based on the agreement of the quadratic model with the actual energy. The global strategy, which has been originally proposed by Fletcher\cite{Fletcher1999} (FLM), guarantees convergence to the closest local minimum for a well-behaved function and can therefore be used to implement a black-box SCF procedure. One of the most attractive features of such a procedure is that it exhibits a quadratic convergence rate, which makes it suited for applications where a very tight convergence of the SCF orbitals and density is required.

Computational cost is, of course, a fundamental aspect that one needs to keep in mind when designing a SCF code. A quadratically convergence SCF (QCSCF) scheme can be implemented in a direct fashion, where all the Hessian-vector products needed to compute the step are performed without assembling the full Hessian matrix and in the atomic orbital (AO) basis. The operations required to perform such a matrix-vector product (MVP) are computationally equivalent to the direct construction of a Fock matrix. Therefore, QCSCF and SCF exhibit the same computational scaling and can benefit both of Cauchy-Schwarz screening in an integral-direct implementation and even of linear-scaling techniques exactly in the same way. Nevertheless, QCSCF requires, in general, a larger number of Fock matrix builds and for well-behaved systems is always going to be more expensive than standard SCF. On the other hand, QCSCF does not require the $\mathcal{O}$($N^3$) diagonalization of the Fock matrix, which makes it in principle advantageous in the asymptotic regime. 

Despite the increased computational cost, many second-order SCF implementations are available. An implementation based on the same algorithm as the one described in the present paper is present in the Dalton suite of programs\cite{daltonpaper} for restricted (RHF) and high-spin restricted open (ROHF) references, and a similar trust-region augmented Hessian implementation has been recently presented by Helmich-Paris\cite{Helmich-Paris2021} and implemented in the ORCA package\cite{Neese2020} for restricted and unrestricted (UHF) references. In the Gaussian 16 suite of programs\cite{g16}, a NR second-order method combined with a linear search when far from the quadratic regions is available for RHF and UHF. Other approaches, either based on a quasi-Newton update\cite{Fischer1992} or on an orbital Hessian based preconditioned conjugate gradient\cite{Wong1995}, can be found in MOLCAS\cite{Karlstrom2003,Aquilante2016} and NWChem\cite{Apra2020}, respectively.

The QCSCF program described in this contribution has been implemented in the CFOUR suite of programs\cite{cfour,Matthews2020}. CFOUR is a quantum chemistry package devoted to high-level post-HF calculations, and therefore the computational cost associated with solving the SCF equations is usually not a main concern for the typical application. Therefore, we do not pursue an integral-direct implementation, even though it would not present any additional difficulty, nor the use of linear-scaling techniques. To achieve some computational efficiency, we offer instead an implementation that can either proceed in a traditional fashion, reading pre-computed two-electron integrals from disk, or use their Cholesky decomposition\cite{Beebe1977,Roeggen1986,Roeggen2008,Koch2003,Weigend2009,Aquilante2011,Folkestad2019} (CD). The latter possibility comes as a part of a long-term goal to deploy the CD machinery for subsequent post-HF calculations\cite{Matthews2020} that has been actively pursued by several developers of the CFOUR suite of programs and that has recently been proposed for complete active space-SCF calculations\cite{Nottoli2021} and for the calculation of NMR chemical shielding tensors at second-order M{\o}ller-Plesset perturbation theory (MP2) using gauge-including atomic orbitals\cite{Burger2021}. On the other hand, having a robust, almost black box SCF implementation is particularly attractive for the users of CFOUR that deal with open-shell systems, where the unrestricted (UHF) and high-spin restricted-open-shell HF (ROHF)\cite{Roothaan1960} SCF equations can be particularly hard to converge. In other words, the main goal of this implementation is to save human time rather than machine time.

The paper is organized as follows. In section \ref{sec:theo}, we briefly recapitulate the norm-extended optimization algorithm and its application to the SCF problem. In section \ref{sec:impl}, we discuss the implementation of the various quantities required for the QCSCF procedure with and without CD of the two-electron integrals. In section \ref{sec:app}, we present a few case studies, which represent prototypical applications of the QCSCF program and that illustrate possible problems and drawbacks that a user can encounter. We finally conclude the paper with a short summary.

\section{Norm-extended optimization of the SCF energy\label{sec:theo}}
In this section, we discuss the main principles of a QCSCF implementation based on the norm-extended optimization (NEO) algorithm. The NEO scheme, originally formulated and implemented by Jensen and coauthors for multiconfigurational SCF wavefunctions\cite{Jensen1983,Jensen1986}, is an elegant and efficient practical realization of the FLM second-order procedure that allows for a direct implementation. In this section, we focus our discussion on the high-spin ROHF optimization problem. In the following, we omit the high-spin specification and refer to the method simply as ROHF. RHF and UHF can be easily derived from the more general ROHF case. The ROHF determinant is parametrized in terms of $N_r$ orbital rotations
\begin{equation}
    \label{eq:ROHFWF}
    |\Phi\rangle = e^{-\kk}|0\rangle,
\end{equation}
where $|0\rangle$ is a reference determinant,
\begin{equation}
\kk = \sum_{ix} \kappa_{ix} \EE_{ix} + \sum_{ia} \kappa_{ia} \EE_{ia} + \sum_{xa}\kappa_{xa}\EE_{xa}
\end{equation}
is the elementary orbital rotation operator, where the rotations mix internal (i, j, ..., doubly occupied) and active (x, y, ..., singly occupied), internal and external (a, b, ..., empty), and active and external orbitals, and $\EE_{pq} = \hat{E}_{pq} - \hat{E}_{qp}$, where $\hat{E}_{pq}$ is a singlet excitation operator (p, q, ..., generic orbitals). The matrix elements of $\kk$ introduce a complete, non-redundant parametrization of orbital rotations that can connect the reference determinant $|0\rangle$ to any non-orthogonal determinant. This fundamental result, known as Thouless' theorem,\cite{Thouless60} is the basis of direct optimization SCF techniques. We define a quadratic model of the SCF energy by expanding the expectation value of the Hamiltonian up to second order in $\kappa$:
\begin{equation}
    \label{eq:QModel}
    \mathcal{Q}(\kappa) = E_0 + \sum_{pq} \kappa_{pq} g_{pq} + \frac{1}{2}\sum_{pqrs} G_{pq,rs}\kappa_{pq}\kappa_{rs},
\end{equation}
where $E_0$ is the reference energy. The orbital-rotation gradient $g \in \mathbb{R}^{N_r}$ is
\begin{equation}
    \label{eq:gradient}
    g_{pq} = \left . \langle 0 | [\EE_{pq},\Ham] | 0\rangle  \right |_{\kappa=0},
\end{equation}
and the orbital-rotation Hessian $G \in \mathbb{R}^{N_r\times N_r}$ is given by
\begin{equation}
    \label{eq:Hessian} G_{pq,rs} = \frac{1}{2}(1+P_{pq,rs})\left . \langle 0 | [\EE_{pq},[\EE_{rs},\Ham]] | 0\rangle  \right |_{\kappa=0},
\end{equation}
where $P_{pq,rs}$ permutes the indices pairs $pq$ and $rs$.
The FLM procedure is an iterative algorithm that computes an optimization step by minimizing the quadratic model in eq.~\ref{eq:QModel} under the constraint that the norm of the step is not larger than a user-defined trust radius $R_t$. This is achieved by introducing the constraint using a Lagrange multiplier $\nu$:
\begin{equation}
    \label{eq:Lag}
    \mathcal{L}(\kappa,\nu) = \mathcal{Q}(\kappa) - \frac{1}{2}\nu (\|\kappa\|^2 - R_t^2).
\end{equation}
The resulting Euler-Lagrange equations, also known as the LM equations, are
\begin{equation}
    \label{eq:LM}
    \left \{
    \begin{array}{l}
    (G - \nu I)\kappa = -g,\\
    \|\kappa\|(\nu) = R_t.
    \end{array}
    \right .
\end{equation}
The trust radius is updated dynamically during the optimization based on the agreement between the energy and its quadratic model. Let $\Delta E$ be the actual energy variation after a LM step, i.e., $\Delta E = E(\kappa) - E_0$ and let $\Delta Q$ be the predicted variation using the quadratic model, i.e., $\Delta Q = Q(\delta) - E_0$, where $\kappa$ is the solution to the LM equations. Let $r = \Delta E/\Delta Q$ be the ratio of the variations. If the ratio is negative, the energy is rising and the step is rejected. The trust radius is reduced by a factor (0.66 in our implementation) and a new step is computed. If the ratio is positive the step is accepted. If $0<r\leq 0.25$, the agreement of the quadratic model with the energy is poor and the trust radius is reduced (again, in our implementation, by a factor 0.66). If $0.25 < r \leq 0.75$ the trust radius is left unchanged, while if $r > 0.75$ the trust radius is increased (by a factor 1.2 in our implementation). The algorithm is robust with respect to the choice of the parameters used to adapt the trust radius, and it can be proven that convergence to a local minimum is always achieved\cite{Fletcher1999}.

In principle, solving the LM equations requires the knowledge of at least the lowest eigenvalue $\lambda_1$ of the Hessian, as it can be proved that, to get to a minimum, the constraint equation needs to be solved in the range $\nu \in (-\infty, \lambda_1)$. In other words, one would need to compute the lowest eigenvalue of the Hessian and then to solve the LM equations. The NEO algorithm is an efficient, combined realization of the two steps that is achieved by introducing a gradient-scaled, augmented Hessian $L(\alpha) \in \mathbb{R}^{N_r+1,N_r+1}$, defined as
\begin{equation}
    \label{eq:NeoL}
    L(\alpha) = \left (
    \begin{array}{cc}
    G & \alpha g \\
    \alpha g^\dagger & 0
    \end{array}
    \right ).
\end{equation}
Let $P$ be an orthogonal projector such that $PL(\alpha)P = G$. The NEO step is given by
\begin{equation}
    \label{eq:NEOStep}
    \kappa = \frac{1}{\alpha s} Py,
\end{equation}
where 
\begin{equation} 
\label{eq:NeoEig}
L(\alpha) y = \nu_1 y,
\end{equation}
i.e., $y$ is the eigenvector of $L(\alpha)$ associated to its lowest eigenvalue, $s = (1-P)y$, and $\alpha$ is obtained by solving numerically the monodimensional equation $\|\kappa\| = R_t$. It can be shown that the NEO step solves the LM equations with a level-shift parameter $\nu = \nu_1$. The Hylleraas-Undheim-MacDonald theorem guarantees that, as $G = PL(\alpha)P$, $\nu_1 \leq \lambda_1$. Therefore, the NEO algorithm converges to a local minimum\cite{Jensen1983}.

The NEO algorithm requires one to compute the lowest eigenvalue of the augmented Hessian $L$, which can be done in a direct fashion using an iterative algorithm such as Davidson diagonalization. It can also be shown that if the vector $(0,\ldots, 0,1)$ is kept into the subspace, it is possible to compute a new step, in case the one computed is rejected, without having to solve the eigenvalue problem in eq.~\ref{eq:NeoEig} again.
As a final note, we remark that, as soon as the optimization has reached a local region and the quadratic approximation becomes valid, the NEO step becomes fully equivalent to a standard NR step. Therefore, in the final stage of the optimization, we switch from solving the NEO equation to the plain NR ones.

\section{Implementation\label{sec:impl}}
Equations for the energy, the gradient, and the MVP can be conveniently obtained by introducing the generalized Fock matrix, $F$, whose elements are obtained as follows
\begin{align}
    \label{eq:genfock}
    &F_{ip} = 2(F^{I}_{ip} + F^{A}_{ip}),\\
    &F_{xp} = Q_{xp} + F^{I}_{xp}, \\
    &F_{ap} = 0,
\end{align}
with $F^{I}_{pq}$, $F^{A}_{pq}$, and $Q_{xp}$ being the inactive Fock matrix, the active Fock matrix, and the Q matrix respectively. The latter can be effectively computed in the AO basis
\begin{align}
    &F^{I}_{\m\n} = h_{\m\n} + \sum_{\rho\s}P^{I}_{\rho\s}\left[(\m\n|\rho\s) - \frac{1}{2}(\m\s|\rho\n)\right],\\
    &F^{A}_{\m\n} = \sum_{\rho\s}P^{A}_{\rho\s}\left[(\m\n|\rho\s) - \frac{1}{2}(\m\s|\rho\n)\right],\\
    &Q_{\m\n} = \sum_{\rho\s}P^{A}_{\rho\s}\left[(\m\n|\rho\s) - (\m\s|\rho\n)\right],
\end{align}
where we have used Mulliken notation for the two-electron integrals and Greek indices to refer to the AOs.
Here, $P^{I}_{\m\n} = 2\sum_{i}C_{\m i}C_{\n i}$ and $P^{A}_{\m\n}=\sum_{u}C_{\m u}C_{\n u}$ are the inactive and active one-body density matrices written in the AO basis, respectively. As it is evident from the equations above, the actual implementation requires minor modifications to the customary routine that assembles the RHF Fock matrix. 

The generalized Fock matrix is transformed into the MO basis and then used to calculate the ROHF energy
\begin{equation}
    E_{\rm ROHF} = \sum_{i}\left(h_{ii} + \frac{1}{2}F_{ii}\right) + \frac{1}{2}\sum_{u}\left(h_{uu}+F_{uu}\right).
\end{equation}
Furthermore, its anti-symmetric part is used to compute the gradient as follows
\begin{equation}
    g_{pq} = 2(F_{pq}-F_{qp}),
\end{equation}
where the only relevant rotations are the ones mixing orbitals belonging to different classes (i.e., internal, active, external).

The eigenvalue problem in eq. \ref{eq:NeoEig} is solved via Davidson diagonalization, while the NR linear system is solved using an iterative preconditioned conjugate-gradient (PCG) solver. Both algorithms are implemented in a matrix-free spirit, that is, they only require one to perform MVPs, and not to build and store in memory the full matrix. 
It is important to stress that the overall algorithm works in the MO basis, as this allows us to exploit the diagonal dominant character of the MO rotation Hessian. 
In the MO basis, the MVP can be written as
\begin{equation}
    \label{eq:MVP}
    b_{pq} = 2(\tilde{F}_{pq}-\tilde{F}_{qp}) + \frac{1}{2}\sum_{r}\left(\kappa_{pr}g_{rq} - g_{pr}\kappa_{qr}\right),
\end{equation}
where $\kappa$ is a trial vector in Davidson's algorithm and where we introduce the one-index transformed generalized Fock matrix $\tilde{F}$, which is defined as in eq. \ref{eq:genfock} but using intermediate matrices computed with density matrices dressed with the trial vector. To avoid transforming the two-electron integrals into the MO basis, we compute the MVP in the AO basis. In particular, we define transformed and symmetrized one-body density matrices:
\begin{align}
    &\tilde{P}^{I}_{\m\n} = 2\sum_{iq}C_{\m i}\kappa_{iq}C_{\n q} + 2\sum_{iq}C_{\n q}\kappa_{qi}C_{\s i},\\
    &\tilde{P}^{A}_{\m\n} = \sum_{uq}C_{\m u}\kappa_{uq}C_{\n q} + \sum_{uq}C_{\n q}\kappa_{qu}C_{\s u},
\end{align}
and use them to assemble the one-index transformed internal and active Fock matrices and Q matrix. These are in turn used to build the one-index transformed generalized Fock matrix, which is then finally transformed back to the MO basis. Therefore, computing the required MVP exhibits a computational cost similar with the one of a standard SCF iteration and, more importantly, the same scaling with respect to the system's size. Specifically, the leading term operation in both the SCF and QCSCF algorithms scale as $\mathcal{O}(N^{4})$, where N is the number of basis functions. However, each QCSCF iteration requires the solution of either the NEO or the NR equations with an iterative solver thus increasing the prefactor of a QCSCF calculation with respect to the standard SCF one.  

When using the CD, the two-electron integrals matrix -- written in the AO basis -- is approximated as follows
\begin{equation}
    (\mu\nu|\rho\sigma) \simeq \sum_{K}^{N_{\rm ch}} L^{K}_{\mu\nu}L^{K}_{\rho\sigma},
\end{equation}
where $L^{K}_{\mu\nu}$ is a Cholesky vector and $N_{\rm ch}$ is the rank of the decomposition.
In this framework, it is convenient to compute the Coulomb and exchange contributions to the various matrices separately\cite{Aquilante2007,Aquilante2011}. 
The Coulomb contribution can be computed by performing a straightforward contraction of a Cholesky vector with the one-body density matrix and then multiplying the resulting factor by a Cholesky vector. On the other hand, in order to compute the exchange contributions, it is convenient to first half-transform the Cholesky vectors into the MO basis, and then assemble the exchange contribution. Considering the inactive Fock matrix as an example we have
\begin{equation}
    F^{I}_{\m\n} = h_{\m\n} + \sum_{K}^{N_{\rm ch}}\left(\sum_{\rho\s}P^{I}_{\rho\s}L^{K}_{\m\n}L^{K}_{\rho\s}
    - \sum_{i}L^{K}_{\m i}L^{K}_{\n i}\right)
\end{equation}
where, $L^{K}_{\m i} = \sum_{\n}C_{\n i}L^{K}_{\m\n}$.
A similar procedure is applied also for the calculation of the one-index transformed matrices.

The restricted and unrestricted SCF implementations can be trivially obtained as sub-cases of the ROHF one. Second-order RHF is simply derived by neglecting the contributions of active orbitals, that is, by setting the active density matrix to zero. Under these circumstances, only the inactive Fock matrix has to be considered. Regarding the second-order implementation of UHF, we have to take into account two different bases -- one for the alpha and one for the beta electrons. As a result, we have two different set of orbital rotation parameters
\begin{equation}
    \kk = \sum_{ia}\kappa^{\alpha}_{ai}\hat{E}^{\alpha}_{ai} 
         + \sum_{ia}\kappa^{\beta}_{ai}\hat{E}^{\beta}_{ai},
\end{equation}
where, $\hat{E}^{\alpha}_{pq}=\hat{a}^{\dagger}_{p\alpha}\hat{a}_{q\alpha}$ and $\hat{E}^{\beta}_{pq}=\hat{a}^{\dagger}_{p\beta}\hat{a}_{q\beta}$. Furthermore, two different density matrices can be obtained namely, $^{\alpha}P^{I}_{\mu\nu}=\sum_{i}C^{\alpha}_{\mu i}C^{\alpha}_{\nu i}$ and $^{\beta}P^{I}_{\mu\nu}=\sum_{i}C^{\beta}_{\mu i}C^{\beta}_{\nu i}$. These are used to assemble the alpha and beta Fock matrices
\begin{align}
    &^{\alpha}F^{I}_{\mu\nu} = h_{\mu\nu} + \sum_{\rho\sigma} {^{\alpha}}P^{I}_{\rho\sigma}
    \left[(\mu\nu|\rho\sigma) - (\mu\sigma|\rho\nu)\right] + 
    \sum_{\rho\sigma}{^{\beta}}P^{I}_{\rho\sigma}(\mu\nu|\rho\sigma),\\
    &^{\beta}F^{I}_{\mu\nu} = h_{\mu\nu} + \sum_{\rho\sigma}{^{\beta}}P^{I}_{\rho\sigma}
    \left[(\mu\nu|\rho\sigma) - (\mu\sigma|\rho\nu)\right] + 
    \sum_{\rho\sigma}{^{\alpha}}P^{I}_{\rho\sigma}(\mu\nu|\rho\sigma).
\end{align}
Accordingly, $^{\alpha}F^{I}$ and $^{\beta}F^{I}$ are used to compute the alpha and beta parts of the gradient respectively. 
A summary of the main equations for the RHF and UHF references can be found in Table \ref{tab:HFrefs}.
\begin{table}[h]
    \centering
    \begin{tabular}{cccc}
         \toprule
         Reference & Energy & Gradient & MVP  \\
         \midrule
         RHF  & $\sum_{i}\left(h_{ii} + F^{I}_{ii}\right)$ 
         & $g_{ai}=-4F^{I}_{ia}$ 
         & $b_{ai} = \tilde{g}_{ai}$\\[5mm]
        \multirow{2}{*}{UHF}& \multirow{2}{*}{$\sum_{i}\left(h_{ii} + \tensor[^{\alpha}]{F}{}^{I}_{ii} + \tensor[^{\beta}]{F}{}^{I}_{ii}\right)$} & 
        $g^{\alpha}_{ai}=-4\tensor[^{\alpha}]{F}{}^{I}_{ia}$ &
        $b^{\alpha}_{ai} = \tilde{g}^{\alpha}_{ai}$\\
        &
        &$g^{\beta}_{ai}=-4\tensor[^{\beta}]{F}{}^{I}_{ia}$&
        $b^{\beta}_{ai} = \tilde{g}^{\beta}_{ai}$\\
         \bottomrule
    \end{tabular}
    \caption{Leading equations used in the second-order implementation of RHF and UHF. We have used a shorthand notation for $\tilde{g}_{ai}$ that means a gradient built with one-index transformed matrices; moreover, this is the only contribution to the MVP since the commutator-like term of eq.~\ref{eq:MVP} vanishes. }
    \label{tab:HFrefs}
\end{table}

We conclude this section with an important remark. Both Davidson diagonalization and the PCG solver require a preconditioner, the most common choice being the inverse diagonal of the matrix. This is a good choice, as the MO rotation Hessian is diagonally dominant in the MO basis. However, assembling the exact diagonal of the Hessian can be expensive. This is not a problem in neither RHF nor UHF, as the leading term of the electronic Hessian's diagonal is given by the difference $F^{I}_{aa}-F^{I}_{ii}$: the two-electron integral contributions to the diagonal can therefore be safely neglected. However, the same is, unfortunately, not the case for ROHF. In particular, the diagonal elements associated with the inactive-active and active-external rotations are poorly approximated by the diagonal elements of the inactive Fock matrices. A much better approximation can be obtained by noting that some of the two-electron contributions to the aforementioned diagonal elements are given by $Q_{ii}$ and $Q_{aa}$. From the theoretical definition of the $Q$ matrix such blocks would not exist; nevertheless, we have direct access to them since we are computing $Q$ in the AO basis. In this way, a good approximation to the Hessian's diagonal can be computed with no additional cost. The inclusion of such terms is fundamental in order to achieve a good convergence for both the Davidson and PCG algorithms. 

\section{Numerical results\label{sec:app}}
In this section, we present numerical results obtained with the quadratic SCF implementation described in the paper. First, we apply the strategy to medium-sized, hard to converge cases to show the numerical stability of the algorithm. Then, we apply the CD version of the code to larger molecules, to show that thanks to the compression operated by the CD technique, calculations can be performed efficiently even for very large systems. Finally, we present a discussion on the possible numerical issues that can arise even when using a quadratically convergent implementation and give some suggestions to rationalize and troubleshoot them. 

The QCSCF algorithm has been implemented in the CFOUR suite of programs\cite{cfour,Matthews2020}, which has been used for all the following QCSCF calculations. In our setup, a few standard SCF iterations are performed to generate a reasonable starting guess for the second-order algorithm. While the second-order optimization can be used for the overall calculation, this is seldom a good idea, as it would require a large number of expensive quadratic steps to reach the local region. On the other hand, a small number of standard SCF iterations are usually enough to get to a good starting point, from which the strong and fast convergence of a second-order scheme can be efficiently exploited.

Specifically, starting from a guess generated by diagonalizing the core-Hamiltonian, we perform up to 30 SCF iterations applying a damping of 0.5 to the first 5 iterations and then using Pulay's DIIS\cite{Pulay1980} to accelerate convergence. We switch to the quadratic algorithm as soon as the root mean square deviation of the density-matrix increment is lower than $10^{-1}$ and its maximum deviation is lower then $1$. 

\subsection{Quadratically convergent calculations on small and large molecules}
\begin{figure}
    \centering
    \includegraphics[width=.9\textwidth]{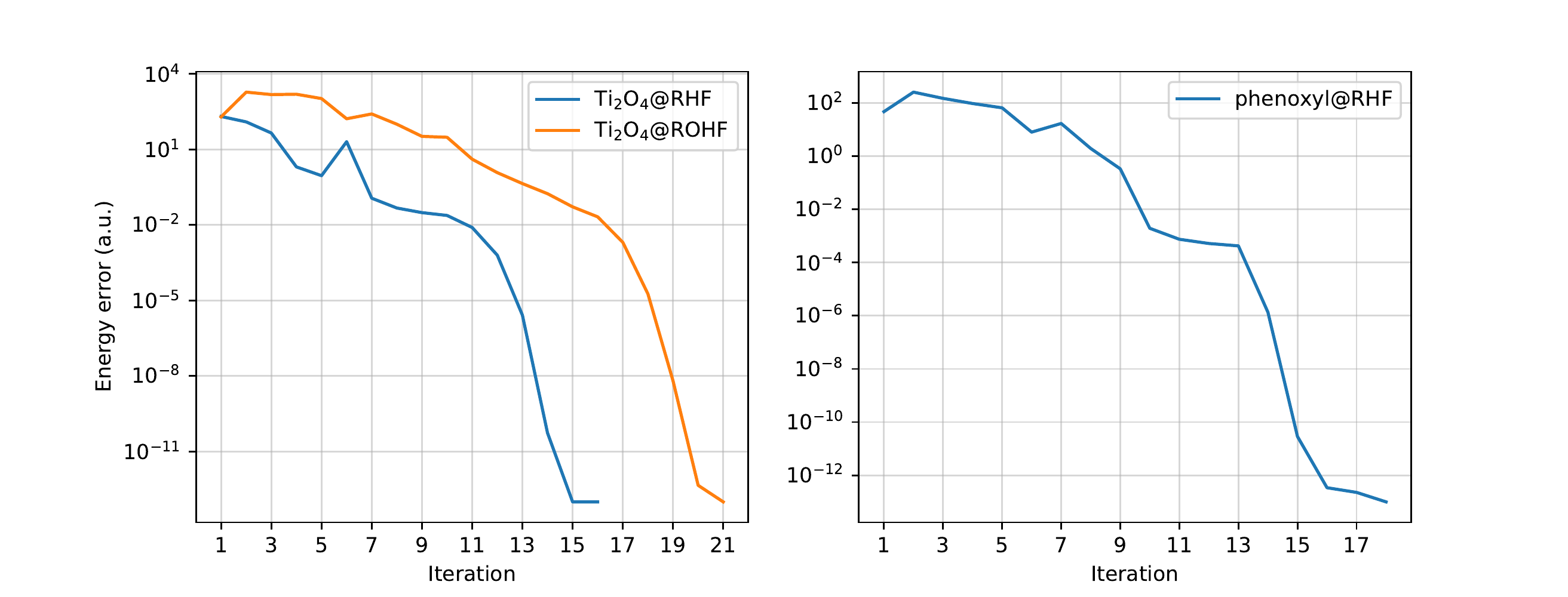}
    \caption{Convergence profile for Ti$_2$O$_4$, both for the RHF (blue) and ROHF (orange) references, and of the phenoxyl radical respectively on the left and right panel. The QC algorithm started at iteration 16 for triplet Ti$_2$O$_4$, 10 for singlet Ti$_2$O$_4$, and 12 for the phenoxyl radical. The final electronic energies for the singlet and triplet Ti$_2$O$_4$ are -1996.144~738~811 and -1996.100~447~937 $E_{h}$ respectively while the one for the phenoxyl radical is -304.953~646~044 $E_{h}$.}
    \label{fig:TiPhe}
\end{figure}

In order to show the black-box nature of the proposed algorithm, we tested it on two systems that exhibit problematic convergence patterns. The first system is $\rm Ti_{2}O_{4}$ in its $ D_{2\textit{h}}$ geometry, where the two titanium atoms lie on one of the three $C_{2}$ axes, the second is the phenoxyl radical. The first system is used as a template to troubleshoot SCF converge issues\footnote{See \url{https://www.scm.com/doc/ADF/Examples/SCF\_Ti2O4.html}} in the Amsterdam Density Functional (ADF) quantum-chemistry package\cite{ADF}. We compute, using Dunning's aug-cc-pVTZ basis set\cite{Kendall1992}, both a singlet and a triplet wavefunction, the latter using a ROHF reference. The second system, a doublet in its ground state, is usually well behaved, but we choose a particular geometry at which two electronic states are nearly intersecting. Again, we use a ROHF reference and we employ Dunning's cc-pVDZ basis set\cite{Dunning1989}. 
We report in Figure \ref{fig:TiPhe} the convergence profile of the calculation on Ti$_2$O$_4$ (left panel) and of the phenoxyl radical (right panel). Both calculations converge reasonably smoothly in a limited number of iterations. It is interesting to note the we could not get the phenoxyl radical ROHF calculation to converge using the standard SCF code in CFOUR using the algorithm described in Ref. \cite{Binkley1974}, despite various attempts using several combinations of DIIS and damping parameters. In other words, using a QCSCF program can turn a labor intensive, possibly fruitless activity into a simple, routine application, at the cost of increased computer time.

\begin{figure}
    \centering
    \includegraphics[width=.7\textwidth]{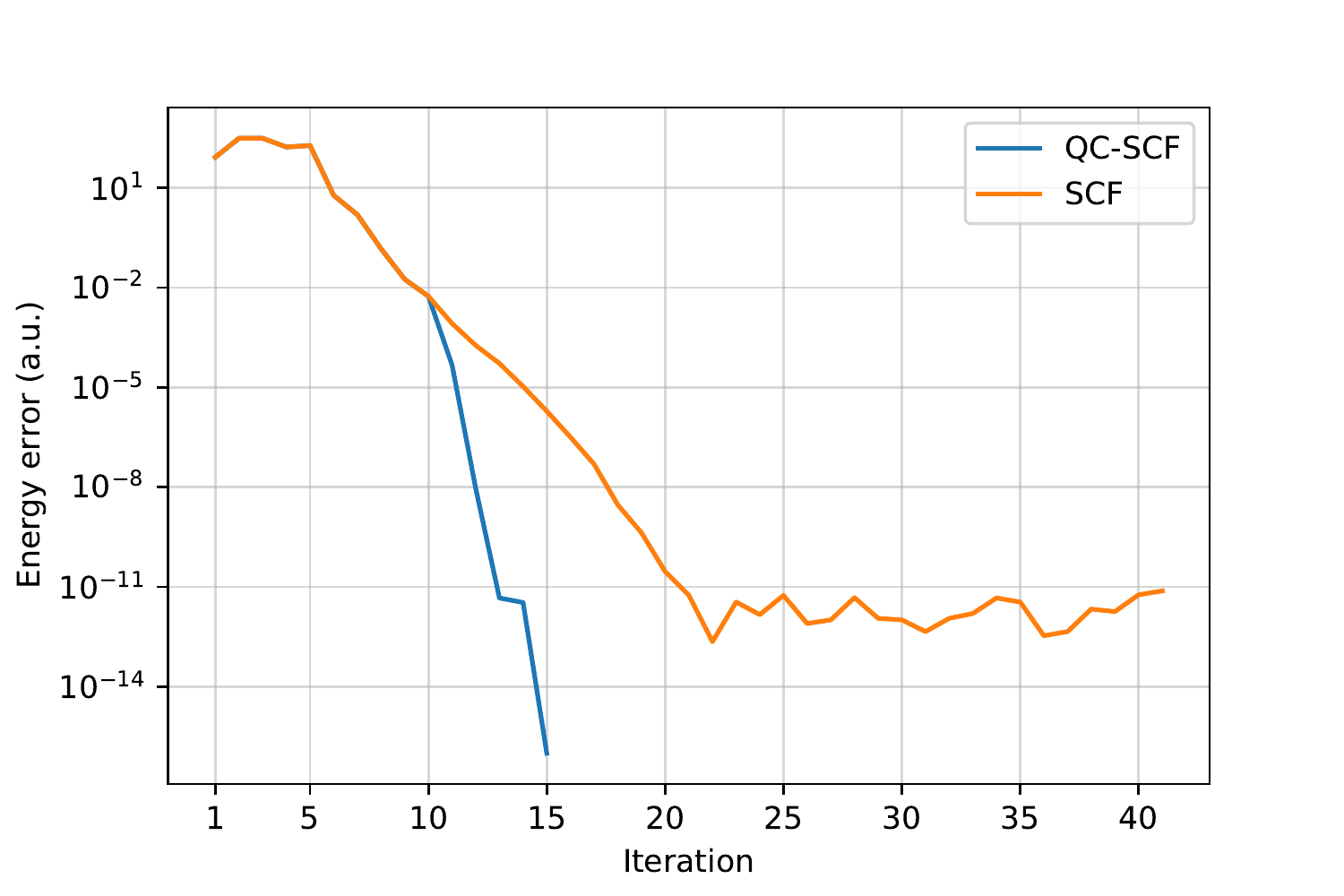}
    \caption{Convergence profile for the PNA molecule for the SCF (orange) and QC-SCF (blue) code using Dunning's aug-cc-pVDZ basis set. The QCSCF code performed 10 regular SCF iterations before switching to the QC algorithm. The final electronic energy is -489.277~111~404 $E_{h}$.}
    \label{fig:pna}
\end{figure}

As a second example of standard use of a QCSCF implementation, we optimize the RHF wavefunction of a small organic molecule, paranitroaniline (PNA), using Dunning's aug-cc-pVDZ basis set\cite{Dunning1989}. This is not a problematic system, as convergence can be easily achieved with a standard SCF code, but becomes an issue for applications where a very tight SCF convergence is required. This is the case when one is interested in computing high-order molecular properties using a post-HF method. A typical example is the calculation of accurate anharmonic force fields, as the ones required for the treatment of anharmonicity. In such applications, cubic and quartic force fields are in general computed by numerically differentiating analytical Hessians, the latter computed for instance using coupled-cluster (CC) theory. To achieve a good numerical accuracy, a tight convergence of the SCF and CC amplitude equations, as well as the various coupled-perturbed equations, is paramount: this can be difficult for a regular SCF code when using extended basis sets. In Figure \ref{fig:pna}, we compare the convergence profile of the standard SCF code in CFOUR with QCSCF. The standard SCF code has no issue achieving a reasonable ($10^{-7}$ to $10^{-8}$ in the RMS norm of the density increment) convergence, but then stagnates and oscillates. On the other hand, the quadratically convergent code achieves effortlessly the required tight convergence ($10^{-11}$ in the RMS norm of the gradient). 

\begin{figure}
    \centering
    \includegraphics[width=.85\textwidth]{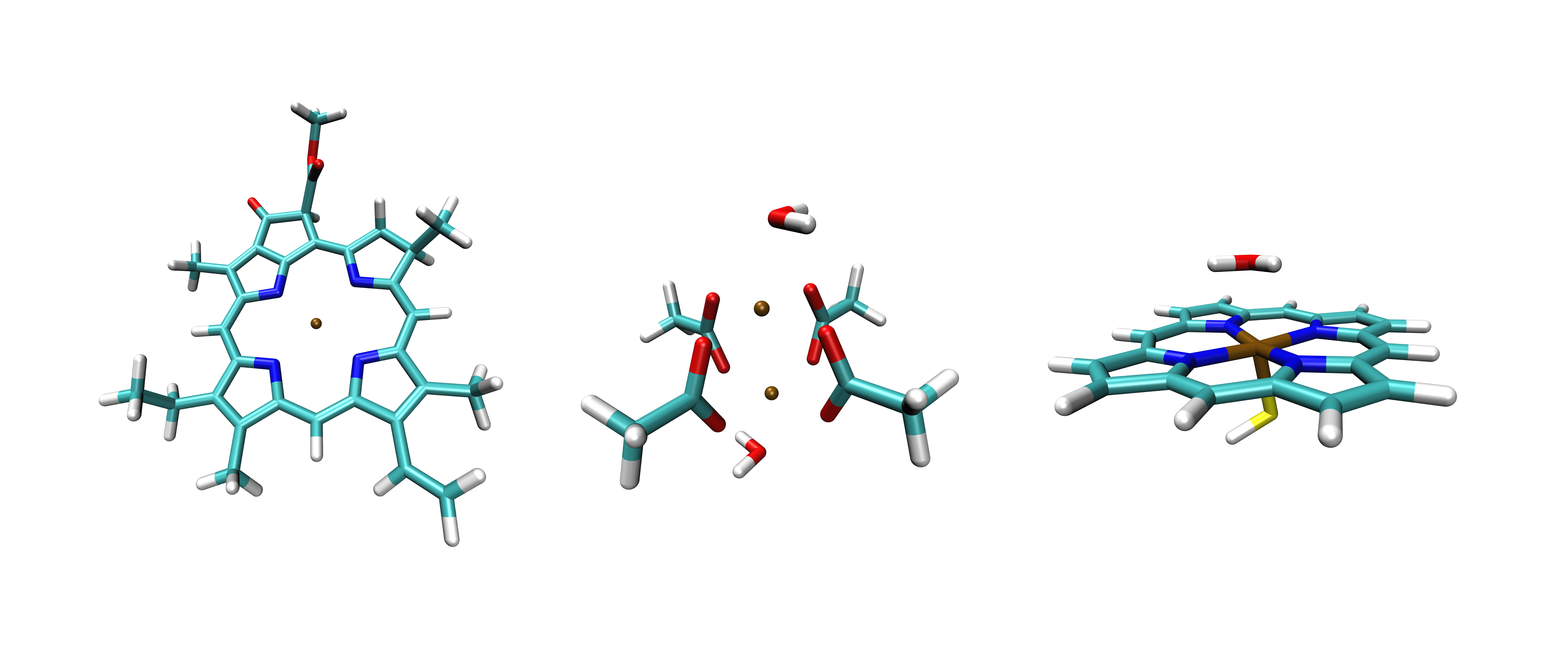}
    \caption{Molecular representations of the three molecules used to test the CD based QCSCF algorithm. From the left: chlorophyll, CUAQUACO2, and HSFe$^{\rm III}$OH$_2$.}
    \label{fig:large}
\end{figure}

The systems proposed so far are medium-sized, and provide examples of applications that are typical for the users of the CFOUR suite of programs. The standard implementation that relies on precomputed two-electron integrals written on disk has been used in all cases. 
\begin{table}[ht]
    \centering
    \begin{tabular}{lcccccc}
        \toprule
         molecule & $^{1}\rm A_{1\textit{g}}$ & It. & $^{3}\rm A_{\textit{u}}$ & It. \\
         \midrule
         $\rm Ti_{2}O_{4}$ & -1996.144~738~811 & 6 & -1996.100~447~937 & 5 & \\
         \bottomrule
    \end{tabular}
    \caption{Results for RHF and ROHF calculation on $\rm Ti_{2}O_{4}$ with the aug-\textit{cc}-pVTZ basis set. The energies are in Hartree units, next to them the number of iterations required by  the second-order optimization. The two calculations were done without the CD and exploiting point-group symmetry.}
    \label{tab:Ti2O4}
\end{table}
As the next examples, we tested the algorithm on three larger systems. Here, the CD of the two-electron integrals has been exploited using a decomposition threshold equal to 10$^{-4}$. The first molecule is an aqua thiolate iron(III) porphyrin complex (HSFe$^{\rm III}$OH$_2$) in its doublet state used as a model system for the active site of the cytochrome P450 in Ref. \cite{Groenhof2005}. The geometry can be found in the Supporting Information of the aforementioned paper. The second calculation was done on the triplet state of a binuclear copper magnet (CUAQUACO2) whose geometry has been taken from Ref. \cite{Feng2019}. Finally, we optimized the singlet state of a chlorophyll molecule where the phytyl tail has been substituted with a hydrogen atom to reduce the computational cost. The geometry was optimized with B3LYP/6-31G(d)\cite{Becke1993,Hehre1972}, using the Gaussian 16\cite{g16} suite of programs. In Figure \ref{fig:large}, a representation of the three systems under study is shown. The optimization of the open-shell systems, i.e., HSFe$^{\rm III}$OH$_2$ and CUAQUACO2, was carried out with ROHF. For all the calculations we adopted the cc-pVTZ basis set\cite{Dunning1989}. In Table \ref{tab:large}, we report the number of basis functions, the number of iterations required to converge the second-order algorithm, the time spent to assemble the Fock matrix ($F^{I}$, $F^{A}$, and $Q$ for the ROHF calculations), and the total CPU wall time for the whole calculation (i.e., preliminary SCF iterations and quadratically convergent ones). All the calculations have been performed on a computer node equipped with two Xeon Gold 5120 CPUs, for a total of 28 cores, and 128 GB of memory. The time spent performing the MVP is not reported since it is similar with the time needed to compute the Fock matrix.
\begin{table}[ht]
    \centering
    \begin{tabular}{lccccc}
    \toprule
         molecule &  N$_b$ &  It. & Fock (s) & Total (min)\\
    \midrule
         HSFe$^{\rm III}$OH$_2$ & 1062 & 13 & 42 & 110.4 \\
         CUAQUACO2 & 900 & 8 & 38 & 45.4 \\
         chlorophyll & 1624  & 6  & 25 & 26.3 \\
    \bottomrule
    \end{tabular}
    \caption{Large systems results. For each of them, we report the number of basis functions, the number of iterations required by the second-order optimization, the time needed to build a Fock matrix (in seconds), and the total CPU wall time for the whole SCF calculation (in minutes).}
    \label{tab:large}
\end{table}

\subsection{What can go wrong and how to deal with that?}
Despite the rigorous and sound convergence properties of a quadratic optimization algorithm, there are still a few issues that can arise in a calculation. First, one must mention that the convergence properties of the FLM algorithm hold in infinite precision. While this is usually not an issue on double precision machines, the effects of finite arithmetic precision become apparent when trying to achieve very tight convergence of the SCF, especially when using very large basis sets comprising diffuse functions. This can be rationalized in terms of overall conditioning of the problem. If the basis set presents near linear dependencies, i.e., if the overlap matrix has small eigenvalues, the numerical precision of the computed quantities degrades. As a rule of thumb, one cannot expect to achieve a convergence tighter than the product of the machine precision times the ratio between the largest and smallest eigenvalues of the overlap matrix. In practice, this means that it is possible to converge the SCF up to 10$^{-11}$--10$^{-12}$ depending on the basis set, which is usually more than sufficient for high precision applications.

That being said, the fact that the QCSCF will converge does not imply that it will always converge to the desired solution. There are two possible, common scenarios that can present. First, the QCSCF code will converge to a state of the same symmetry as its starting point, but not the desired state. For well-behaved systems, the few SCF iterations performed in the beginning are usually enough to establish, using the Aufbau principle, the right occupation. The user should however be wary that there is no guarantee that this will work automatically: it is therefore a good practice to specify explicitly the occupation, or the symmetry of the state, to ensure the convergence of the optimization procedure to the desired wavefunction. 

A second case that requires some attention by the user is, in general, any UHF calculation. It is well known that the solution to the UHF problem is not unique and that many solutions with different levels of spin contamination can exist. A minimization algorithm will converge to the closest local minimum, so there is no guarantee that the QCSCF solution will be the global minimum. It is the experience of the authors that, with respect to a standard SCF code, a QCSCF optimization tends to converge to the solution lowest in energy and with highest spin contamination. After a UHF calculation, the user needs to check whether the obtained solution is acceptable. 

A third, not common, case that can be encountered is convergence to an unstable solution. This is something that typically happens when performing a calculation on a symmetric system without enforcing point-group symmetry and is somewhat similar to what has already been discussed above about convergence to a state with the wrong occupation. The second-order code should in principle always converge to a minimum; however, there are numerical reasons that may leave the optimization stuck into a saddle point. This is a consequence of the parametrization choice, as Thouless' theorem\cite{Thouless60} specifically states that the determinant one can obtain with a rotation parametrized as in eq. \ref{eq:ROHFWF} cannot be orthogonal to the reference. Therefore, if the QCSCF optimization starts from a determinant orthogonal to the minimum, whether due to imposed or to numerical symmetry constraints, it is prevented from reaching the minimum itself. If the minimum has the same symmetry of the starting point -- including if no point-group symmetry is enforced -- this can easily be avoided by perturbing, at the beginning of the second order optimization, the MO gradient with random noise. On the other hand, if symmetry is enforced and there exists a broken-symmetry solution, this cannot be reached with a second-order method. The latter case can be diagnosed doing a stability analysis and resolved, if the broken-symmetry solution is of interest, enforcing symmetry in a lower subgroup or by completely removing the enforcement of symmetry.

We illustrate all the discussed problematic behaviors using the triplet state of \textit{ortho}-benzyne (o-benzyne) as an example. Such a molecule exhibits $C_{2\textit{v}}$ point-group symmetry and has a singlet ground electronic state. This molecule has also been used by Tsuchimochi and Scuseria as a test case for their constrained UHF method\cite{Tsuchimochi2010,Tsuchimochi2011}. We follow their procedure and optimize the geometry using density-functional theory, namely, the B3LYP functional\cite{Becke1993} in conjunction with Pople's 6-31G(d) basis set\cite{Hehre1972}. The geometry optimization has been performed using the Gaussian 16\cite{g16} suite of programs. 
On the optimized geometry, we compute the UHF wavefunction with Dunning's cc-pVTZ basis set\cite{Dunning1989} using four different setups, namely i) enforcing point-group symmetry, but without specifying the occupation numbers; ii) enforcing point-group symmetry and specifying the occupation numbers; iii) without symmetry, using the normal setup; iv) without symmetry, adding a small perturbation (0.01 times the gradient's norm times a uniformly distributed random number between -0.5 and 0.5) to the gradient at the beginning of the optimization.
The results, together with a short comment, are reported in table \ref{tab:o-benz}.
\begin{table}[ht]
    \centering
    \begin{tabular}{lccr}
    \toprule
         Setup & Energy & Spin contamination & Comment \\
    \midrule
        i&-229.261~071 & 0.0204 & wrong occupation \\
        ii&-229.468~767 & 0.0277 & symmetric solution\\
        iii&-229.261~071 & 0.0204 & unstable \\
        iv&-229.471~139 & 0.4156 & symmetry broken\\
    \bottomrule
    \end{tabular}
    \caption{Results of UHF calculations for \textit{o}-benzyne with four different setups. Setup i): point-group symmetry enforced, no initial occupation given. Setup ii): point-group symmetry enforced, initial occupation given in input. Setup iii): no symmetry enforced. Setup iv): no symmetry enforced, the gradient is perturbed at the beginning of the second-order optimization. We report the electronic energy (in Hartree), the spin contamination of the wavefunction, and a short comment that describes the solution found.}
    \label{tab:o-benz}
\end{table}
In the first setup, the initial SCF iterations are tasked with computing the occupation of the wavefunction using the Aufbau principle. When convergence of the preliminary SCF is reached, the resulting electronic configuration is the following: 10 doubly occupied $a_1$ orbitals, 1 doubly and one singly occupied $b_1$ orbitals, 8 doubly occupied $b_2$ orbitals and a singly occupied $a_2$ orbital. QCSCF then converges without problems the given state, which is however not the ground triplet state. In the second setup, we specify in input the correct occupations (9 doubly and one singly occupied $a_1$ orbitals, 2 doubly occupied $b_1$ orbitals, 7 doubly and one singly occupied $b_2$ orbitals and one doubly occupied $a_2$ orbitals), which results in the correct behaviour. We note that, while the states that result from the first two setups both have $B_2$ symmetry, the occupations make them strictly orthogonal. In other words, no symmetry-allowed orbital rotation can link the two states, making therefore it impossible for the QCSCF code to converge from one to the other. 

In the third setup, no symmetry is enforced; however, the initial SCF iterations make the QC procedure start very close to an unstable stationary point, to which the optimization gets stuck. It is interesting to note that the unstable solution found with this setup is exactly the same found with setup i), i.e., enforcing symmetry but without specifying the occupation. This suggests that the QC optimizer started from a wavefunction that is orthogonal to the actual minimum. In other words, even though symmetry was not enforced, it was still present numerically, which explains the observed behavior.

In the fourth setup, we repeated the calculation without symmetry, but added a small, random perturbation to the gradient at the beginning of the QCSCF process. The small amount of random noise allowed the QCSCF optimizer to connect to the orthogonal subspace where the minimum lies and converge to it without particular effort. It is interesting to note that the optimization found a solution that has a lower energy than the symmetric one, and a much larger spin contamination. This is the same solution found by Tsuchimochi and Scuseria. A comparison between the UHF solutions found with the various setups is particularly interesting to understand the behavior of the QCSCF optimization. Performing a stability analysis, the symmetric solution from the second setup is found to exhibit an instability towards a symmetry-broken UHF solution, which is exactly the one found with the fourth setup. The latter is in turn stable. All these results confirm numerically what was discussed above. In setups i) and iii), QCSCF converges to a solution with the wrong occupation -- which is explicitly enforced in the first case, and is a numerical consequence of the preliminary SCF iterations in the third. In the second setup, the minimum within the symmetry is found. In the fourth setup, as no symmetry is enforced and the symmetry is artificially broken with a small, random perturbation, a true minimum is found, which corresponds to the stable, symmetry broken solution. 
It is worth commenting that the lowest energy, stable solution exhibits a remarkably large spin contamination, besides being symmetry broken: whether this is the solution of interest, or whether the symmetry one found with the second setup is preferable, is ultimately a choice left to the user. 

We conclude this section with some considerations concerning computational efficiency. The choice of parameters that we adopt by default, namely, the parameters that control the preliminary SCF iterations and the initial trust radius, represent a good compromise between robustness and efficiency. This is, of course, not something that applies in general to every molecular system. It is therefore likely that there exists, for any system, a specific combination of parameters that minimizes the number of overall iterations, and therefore the computational cost. Tuning these parameters is not required to achieve convergence, but may speed up the computation, which is particularly relevant for applications involving large and very large systems. Nevertheless, the spirit of a quadratically convergent SCF is to minimize human effort with respect to computational effort: it is up to the user to decide whether to invest time into the numerical optimization of the procedure for a specific application. Let us remark once again that, even if an optimal setup is found, a QCSCF calculation is inevitably going to be more expensive, heuristically, about twice to three times as expensive, than a well-behaving regular SCF one. 

\section{Conclusions}
In this contribution, we presented a quadratically convergent self-consistent field program that can achieve convergence of the SCF iteration in a black-box manner for restricted, unrestricted, and (high-spin) restricted-open-shell references. The implementation is based on the Fletcher-Levenberg-Marquardt trust-radius Newton method, in its direct formulation known as norm-extended optimization. All the operation are performed in a direct fashion, without the need of assembling explicitly the MO rotation Hessian nor of transforming the two-electron integrals into the MO basis. To the best of our knowledge, this is the first QCSCF implementation with guaranteed convergence that can be applied to both ROHF and UHF.
The resulting algorithm is suited to be used in conjunction with integral-direct, and even linear-scaling techniques to increase its computational performances. As this implementation is meant for high-accuracy applications on small- to medium-sized molecules, we explore a different strategy to reduce computational cost, namely, the use of Cholesky decomposition of the two-electron integrals. 
We showed with numerical examples how the QCSCF code can be helpful not only with hard-to-converge cases, but also with regular cases where a very tight convergence is needed. Using the CD of the two-electron integrals, larger systems become easily treatable.

While a QCSCF calculation is in general more expensive than its linear counterpart, we showed that it is possible to achieve the same scaling with respect to the size of the system as in regular SCF. 
Furthermore, as regular and QC SCF share as the leading computational operation the construction of Fock matrices, they can exploit the same techniques to accelerate the computation, including integral direct implementations, linear scaling methods and low-rank approximations of the two-electron integrals matrix. As an example, we showed an implementation based on the Cholesky decomposition of the two-electron integrals, as the use of such a technique is part of a widespread effort amongst the developers of the CFOUR suite of programs.

A second-order optimization algorithm has the remarkable property of being completely predictable. Its black-box nature is shown in its ability to always converge to a solution, which, however may not be the desired one. We discussed and rationalized the main reasons why QCSCF can converge to either a solution with the wrong symmetry or to an unstable solution, and showed how one can overcome such difficulties by either guiding the optimizer to the right symmetry state by specifying an occupation or, when working enforcing either no or a reduced symmetry point-group, by adding to the gradient with a small random perturbation. 

In conclusion, a second-order SCF code can be a useful tool to converge problematic cases to a very tight threshold in an almost automated way, without the need of tweaking and tuning various parameters to achieve the desired accuracy. We hope that the QCSCF program that will be made available in the next release of CFOUR will provide the community with a useful tool.

\section*{Acknowledgments} This work is dedicated to Prof. John F. Stanton -- quite possibly the biggest fan of the QCSCF code in CFOUR -- on the occasion of his 60th birthday.

\bibliography{qcscf}

\begin{thebibliography}{10}
\providecommand{\url}[1]{\normalfont{#1}}
\providecommand{\urlprefix}{}

\bibitem{Kohn65}
Kohn, W.; Sham, L.J. Self-consistent equations including exchange and
  correlation effects, \emph{Phys. Rev.}  \textbf{1965}, \emph{140},
  A1133--A1138.

\bibitem{Hartree1957}
Hartree, D.R. \emph{{The calculation of atomic structures}}; Wiley, London,
  1957.

\bibitem{Thiel2000}
Thiel, W. \emph{Semiempirical Methods}, Grotendorst, J., Ed.; John von Neumann
  Institute, Jülich, 2000; pp 261--283.

\bibitem{Roothan51}
Roothaan, C.C.J. New developments in molecular orbital theory, \emph{Rev. Mod.
  Phys.}  \textbf{1951}, \emph{23}, 69--89.

\bibitem{Roothaan1960}
Roothaan, C.C. {Self-consistent field theory for open shells of electronic
  systems}, \emph{Rev. of Mod. Phys.}  \textbf{1960}, \emph{32}, 179--185.

\bibitem{Rabuck1999}
Rabuck, A.D.; Scuseria, G.E. {Improving self-consistent field convergence by
  varying occupation numbers}, \emph{J. Chem. Phys.}  \textbf{1999},
  \emph{110}, 695--700.

\bibitem{Cances2000}
Cancès, E. \emph{SCF algorithms for HF electronic calculations}; Springer,
  Berlin, Heidelberg, 2000; pp 17--43.

\bibitem{Pulay1980}
Pulay, P. {Convergence acceleration of iterative sequences - the case of SCF
  iteration}, \emph{Chem. Phys. Lett.}  \textbf{1980}, \emph{73}, 393--398.

\bibitem{Pulay1982}
Pulay, P. {Improved SCF convergence acceleration}, \emph{J. Comput. Chem.}
  \textbf{1982}, \emph{3}, 556--560.

\bibitem{Hamilton1985}
Hamilton, T.P.; Pulay, P. {Direct inversion in the iterative subspace (DIIS)
  optimization of open-shell, excited-state, and small multiconfiguration SCF
  wave functions}, \emph{J. Chem. Phys.}  \textbf{1985}, \emph{84}, 5728--5734.

\bibitem{Sellers1993}
Sellers, H. {The C$^2$-DIIS Convergence Acceleration Algorithm}, \emph{Int. J.
  Quant. Chem.}  \textbf{1993}, \emph{45}, 31--41.

\bibitem{Kudin2002}
Kudin, K.N.; Scuseria, G.E.; Canc{\`{e}}s, E. {A black-box self-consistent
  field convergence algorithm: One step closer}, \emph{J. Chem. Phys.}
  \textbf{2002}, \emph{116}, 8255--8261.

\bibitem{Karlstrom1979}
Karlstr{\"{o}}m, G. {Dynamical damping based on energy minimization for use ab
  initio molecular orbital SCF calculations}, \emph{Chem. Phys. Lett.}
  \textbf{1979}, \emph{67}, 348--350.

\bibitem{Saunders1973}
Saunders, V.R.; Hillier, I.H. {A “Level–Shifting” method for converging
  closed shell Hartree–Fock wave functions}, \emph{Int. J. Quant. Chem.}
  \textbf{1973}, \emph{7}, 699--705.

\bibitem{Vacek1999}
Vacek, G.; Perry, J.K.; Langlois, J.M. {Advanced initial-guess algorithm for
  self-consistent-field calculations on organometallic systems}, \emph{Chem.
  Phys. Lett.}  \textbf{1999}, \emph{310}, 189--194.

\bibitem{Lethola19}
Lehtola, S. Assessment of initial guesses for self-consistent field
  calculations. Superposition of atomic potentials: Simple yet efficient,
  \emph{J. Chem. Theory Comput.}  \textbf{2019}, \emph{15}, 1593--1604.

\bibitem{Bacskay1981}
Bacskay, G.B. {A quadratically convergent Hartree-Fock (QC-SCF) method.
  Application to closed shell systems}, \emph{Chem. Phys.}  \textbf{1981},
  \emph{61}, 385--404.

\bibitem{Bacskay1982}
Bacskay, G.B. {A quadratically convergent Hartree-Fock (QC-SCF) method.
  Application to open shell orbital optimization and coupled perturbed
  Hartree-Fock calculations}, \emph{Chem. Phys.}  \textbf{1982}, \emph{65},
  383--396.

\bibitem{Fletcher1999}
Fletcher, R. \emph{{Practical Methods of Optimization}}, 2nd ed.; Wiley: New
  York, 1999.

\bibitem{daltonpaper}
Aidas, K.; Angeli, C.; Bak, K.L.; Bakken, V.; Bast, R.; Boman, L.;
  Christiansen, O.; Cimiraglia, R.; Coriani, S.; Dahle, P.; et~al. {The Dalton
  quantum chemistry program system}, \emph{WIREs Comput.~Mol.~Sci.}
  \textbf{2014}, \emph{4}, 269--284.

\bibitem{Helmich-Paris2021}
Helmich-Paris, B. {A trust-region augmented Hessian implementation for
  restricted and unrestricted Hartree--Fock and Kohn--Sham methods}, \emph{J.
  Chem. Phys.}  \textbf{2021}, \emph{154}, 164104.

\bibitem{Neese2020}
Neese, F.; Wennmohs, F.; Becker, U.; Riplinger, C. {The ORCA quantum chemistry
  program package}, \emph{J. Chem. Phys.}  \textbf{2020}, \emph{152}, 224108.

\bibitem{g16}
Frisch, M.J.; Trucks, G.W.; Schlegel, H.B.; Scuseria, G.E.; Robb, M.A.;
  Cheeseman, J.R.; Scalmani, G.; Barone, V.; Petersson, G.A.; Nakatsuji, H.;
  et~al. Gaussian~16 {R}evision {A}.03, 2016. Gaussian Inc. Wallingford CT.

\bibitem{Fischer1992}
Fischer, T.H.; Alml{\"{o}}f, J. {General methods for geometry and wave function
  optimization}, \emph{J. Phys. Chem.}  \textbf{1992}, \emph{96}, 9768--9774.

\bibitem{Wong1995}
Wong, A.T.; Harrison, R.J. {Approaches to large‐scale parallel
  self‐consistent field calculations}, \emph{J. Comput. Chem.}
  \textbf{1995}, \emph{16}, 1291--1300.

\bibitem{Karlstrom2003}
Karlstr{\"{o}}m, G.; Lindh, R.; Malmqvist, P.{\AA}.; Roos, B.O.; Ryde, U.;
  Veryazov, V.; Widmark, P.O.; Cossi, M.; Schimmelpfennig, B.; Neogrady, P.;
  et~al. {MOLCAS: A program package for computational chemistry}, \emph{Comput.
  Mat. Sci.}  \textbf{2003}, \emph{28}, 222--239.

\bibitem{Aquilante2016}
Aquilante, F.; Autschbach, J.; Carlson, R.K.; Chibotaru, L.F.; Delcey, M.G.;
  {De Vico}, L.; {Fdez. Galv{\'{a}}n}, I.; Ferr{\'{e}}, N.; Frutos, L.M.;
  Gagliardi, L.; et~al. {MOLCAS 8: New capabilities for multiconfigurational
  quantum chemical calculations across the periodic table}, \emph{J. Comput.
  Chem.}  \textbf{2016}, \emph{37}, 506--541.

\bibitem{Apra2020}
Apr{\`{a}}, E.; Bylaska, E.J.; {De Jong}, W.A.; Govind, N.; Kowalski, K.;
  Straatsma, T.P.; Valiev, M.; {Van Dam}, H.J.; Alexeev, Y.; Anchell, J.;
  et~al. {NWChem: Past, present, and future}, \emph{J. Chem. Phys.}
  \textbf{2020}, \emph{152}, 184102.

\bibitem{cfour}
Stanton, J.F.; Gauss, J.; Cheng, L.; Harding, M.E.; Matthews, D.A.; Szalay,
  P.G. {CFOUR, Coupled-Cluster techniques for Computational Chemistry, a
  quantum-chemical program package}. {W}ith contributions from {A}.{A}. {A}uer,
  {R}.{J}. {B}artlett, {U}. {B}enedikt, {C}. {B}erger, {D}.{E}. {B}ernholdt,
  {S.} {B}laschke, {Y}. {J}. {B}omble, {S.} {B}urger, {O}. {C}hristiansen, {D.}
  Datta, {F}. Engel, {R}. Faber, {J.} {G}reiner, {M}. {H}eckert, {O}. {H}eun,
  {M}. Hilgenberg, {C}. {H}uber, {T}.-{C}. {J}agau, {D}. {J}onsson, {J}.
  {J}us{\'e}lius, {T}. Kirsch, {K}. {K}lein, {G}.{M.} Kopper, {W}.{J}.
  {L}auderdale, {F}. {L}ipparini, {T}. {M}etzroth, {L}.{A}. {M}{\"u}ck,
  {D}.{P}. {O}'{N}eill, {T.} {N}ottoli, {D}.{R}. {P}rice, {E}. {P}rochnow, {C}.
  {P}uzzarini, {K}. {R}uud, {F}. {S}chiffmann, {W}. {S}chwalbach, {C}.
  {S}immons, {S}. {S}topkowicz, {A}. {T}ajti, {J}. {V}{\'a}zquez, {F}. {W}ang,
  {J}.{D}. {W}atts and the integral packages {MOLECULE} ({J}. {A}lml{\"o}f and
  {P}.{R}. {T}aylor), {PROPS} ({P}.{R}. {T}aylor), {ABACUS} ({T}. {H}elgaker,
  {H}.{J}. {A}a. {J}ensen, {P}. {J}{\o}rgensen, and {J}. {O}lsen), and {ECP}
  routines by {A}. {V}. {M}itin and {C}. van {W}{\"u}llen. {F}or the current
  version, see http://www.cfour.de, Last accessed 15 November 2020.

\bibitem{Matthews2020}
Matthews, D.A.; Cheng, L.; Harding, M.E.; Lipparini, F.; Stopkowicz, S.; Jagau,
  T.C.; Szalay, P.G.; Gauss, J.; Stanton, J.F. {Coupled-cluster techniques for
  computational chemistry: The CFOUR program package}, \emph{J. Chem. Phys.}
  \textbf{2020}, \emph{152}, 214108.

\bibitem{Beebe1977}
Beebe, N.H.F.; Linderberg, J. {Simplifications in the generation and
  transformation of two-electron integrals in molecular calculations},
  \emph{Int. J. Quantum Chem.}  \textbf{1977}, \emph{12}, 683--705.

\bibitem{Roeggen1986}
R{\o}eggen, I.; Wisl{\o}ff-Nilssen, E. {On the Beebe-Linderberg two-electron
  integral approximation}, \emph{Chem. Phys. Lett.}  \textbf{1986}, \emph{132},
  154--160.

\bibitem{Roeggen2008}
R{\o}eggen, I.; Johansen, T. {Cholesky decomposition of the two-electron
  integral matrix in electronic structure calculations}, \emph{J. Chem. Phys.}
  \textbf{2008}, \emph{128}, 194107.

\bibitem{Koch2003}
Koch, H.; {S{\'{a}}nchez De Mer{\'{a}}s}, A.; Pedersen, T.B. {Reduced scaling
  in electronic structure calculations using Cholesky decompositions}, \emph{J.
  Chem. Phys.}  \textbf{2003}, \emph{118}, 9481--9484.

\bibitem{Weigend2009}
Weigend, F.; Kattannek, M.; Ahlrichs, R. {Approximated electron repulsion
  integrals: Cholesky decomposition versus resolution of the identity methods},
  \emph{J. Chem. Phys.}  \textbf{2009}, \emph{130}, 164106.

\bibitem{Aquilante2011}
Aquilante, F.; Boman, L.; Bostr{\"o}m, J.; Koch, H.; Lindh, R.; de~Mer{\'a}s,
  A.S.; Pedersen, T.B. \emph{Cholesky decomposition techniques in electronic
  structure theory}, Zalesny, R., Papadopoulos, M.G., Mezey, P.G., Leszczynski,
  J., Eds.; Springer, Dordrecht, 2011; Vol.~13, pp 301--343.

\bibitem{Folkestad2019}
Folkestad, S.D.; Kj{\o}nstad, E.F.; Koch, H. {An efficient algorithm for
  Cholesky decomposition of electron repulsion integrals}, \emph{J. Chem.
  Phys.}  \textbf{2019}, \emph{150}, 194112.

\bibitem{Nottoli2021}
Nottoli, T.; Gauss, J.; Lipparini, F. A Second-order {CASSCF} algorithm with
  the {C}holesky decomposition of the two-electron repulsion integrals,
  \emph{J. Chem. Theory Comput.}  \textbf{2021}, \emph{xx}, submitted.

\bibitem{Burger2021}
Burger, S.; Lipparini, F.; Gauss, J.; Stopkowicz, S. {NMR} chemical shift
  computations at second-order {M}{\o}ller-{P}lesset perturbation theory using
  gauge-including atomic orbitals and {C}holesky-decomposed two-electron
  integrals, \emph{J. Chem. Phys.}  \textbf{2021}, \emph{xx}, submitted.

\bibitem{Jensen1983}
Jensen, H.J.{\relax Aa}.; J{\o}rgensen, P. {A direct approach to second‐order
  MCSCF calculations using a norm extended optimization scheme}, \emph{J. Chem.
  Phys.}  \textbf{1984}, \emph{80}, 1204--1214.

\bibitem{Jensen1986}
Jensen, H.J.{\relax Aa}.; {\r A}gren, H. {A Direct, restricted-step,
  second-order MC SCF program for large scale ab initio calculations},
  \emph{Chem. Phys.}  \textbf{1986}, \emph{104}, 229--250.

\bibitem{Thouless60}
Thouless, D. Stability conditions and nuclear rotations in the Hartree-Fock
  theory, \emph{Nuc.Phys.}  \textbf{1960}, \emph{21}, 225--232.

\bibitem{Aquilante2007}
Aquilante, F.; Pedersen, T.B.; Lindh, R. {Low-cost evaluation of the exchange
  Fock matrix from Cholesky and density fitting representations of the electron
  repulsion integrals}, \emph{J. Chem. Phys.}  \textbf{2007}, \emph{126},
  194106.

\bibitem{ADF}
te~Velde, G.; Bickelhaupt, F.M.; Baerends, E.J.; Fonseca~Guerra, C.; van
  Gisbergen, S.J.A.; Snijders, J.G.; Ziegler, T. Chemistry with {ADF}, \emph{J.
  Comput. Chem.}  \textbf{2001}, \emph{22}, 931--967.
  \urlprefix\url{http://dx.doi.org/10.1002/jcc.1056}.

\bibitem{Kendall1992}
Kendall, R.A.; Dunning, T.H.; Harrison, R.J. {Electron affinities of the
  first-row atoms revisited. Systematic basis sets and wave functions},
  \emph{J. Chem. Phys.}  \textbf{1992}, \emph{96}, 6796--6806.

\bibitem{Dunning1989}
{Dunning, Jr.}, T.H. {Gaussian basis sets for use in correlated molecular
  calculations. I. The atoms boron through neon and hydrogen}, \emph{J. Chem.
  Phys.}  \textbf{1989}, \emph{1007}, 4572--4585.

\bibitem{Binkley1974}
Binkley, J.; Pople, J.A.; Dobosh, P.A. {The calculation of spin-restricted
  single-determinant wavefunctions}, \emph{Mol. Phys.}  \textbf{1974},
  \emph{28}, 1423--1429.

\bibitem{Groenhof2005}
Groenhof, A.R.; Swart, M.; L; Ehlers, A.W.; Lammertsma, K. {Electronic ground
  states of iron porphyrin and of the first species in the catalytic reaction
  cycle of cytochrome P450s}, \emph{J. Phys. Chem. A}  \textbf{2005},
  \emph{109}, 3411--3417.

\bibitem{Feng2019}
Feng, X.; Epifanovsky, E.; Gauss, J.; Krylov, A.I. {Implementation of analytic
  gradients for CCSD and EOM-CCSD using Cholesky decomposition of the
  electron-repulsion integrals and their derivatives: Theory and benchmarks},
  \emph{J. Chem. Phys.}  \textbf{2019}, \emph{151}, 014110.

\bibitem{Becke1993}
Becke, A.D. {A new mixing of Hartree-Fock and local density-functional
  theories}, \emph{J. Chem. Phys.}  \textbf{1993}, \emph{98}, 1372--1377.

\bibitem{Hehre1972}
Hehre, W.J.; Ditchfield, R.; Pople, J.A. {Self — Consistent molecular orbital
  methods . XII . Further extensions of Gaussian — type basis sets for use in
  molecular orbital studies of organic molecules}, \emph{J. Chem. Phys.}
  \textbf{1972}, \emph{56}, 2257--2261.

\bibitem{Tsuchimochi2010}
Tsuchimochi, T.; Scuseria, G.E. {Communication: ROHF theory made simple},
  \emph{J. Chem. Phys.}  \textbf{2010}, \emph{133}, 141102.

\bibitem{Tsuchimochi2011}
Tsuchimochi, T.; Scuseria, G.E. {Constrained active space unrestricted
  mean-field methods for controlling spin-contamination}, \emph{J. Chem. Phys.}
   \textbf{2011}, \emph{134}, 064101.

\end{thebibliography}

\end{document}